\documentclass{article}
\usepackage[english]{babel}
\usepackage{amsmath}

\usepackage{amsfonts}
\usepackage{amssymb}
\usepackage{bm}
\usepackage[pdftex]{graphicx}
\usepackage{setspace}
\usepackage{color}
\usepackage{hyperref}
\usepackage{fancyhdr}
\begin{document}

\begin{center}
{\LARGE \textbf{Interplay between Phonon Confinement and Fano Effect on Raman line shape for semiconductor nanostructures: Analytical study}}
\end{center}

\begin{center}

\end{center}
\textbf{Priyanka Yogi$^{1,\$}$, Shailendra K. Saxena$^{1, \$}$, Suryakant Mishra$^{1}$, Vikash Mishra$^{1}$, Hari M. Rai$^{1}$, Ravikiran Late$^{1}$, Vivek Kumar$^{2}$, Bipin Joshi$^{3}$, Pankaj R. Sagdeo$^{1}$ and Rajesh Kumar$^{1, *}$}

\vspace{0.5cm}

\textit{$^1$ Material Research Laboratory, Discipline of Physics, IIT Indore,Simrol-452020, Madhya Pradesh, India}

\textit{$^2$ Department of Physics, National Institute of Technology Meghalaya, Shillong-793003, Meghalaya, India}

\textit{$^3$ Department of Science and Technology (DST), Technology Bhavan,New Delhi, Delhi 110016}

\vspace{0.2cm}

$^{\$}$ Authors having equal contribution 

\vspace{0.2cm}

* Corresponding author 
email: rajeshkumar@iiti.ac.in 

\vspace{0.5cm}

KEYWORDS: Raman Line-Shape; Electron-Phonon Interaction, Fano Effect

\vspace{0.5cm}

\subsection*{Abstract} Theoretical Raman line shape functions have been studied to take care of quantum confinement effect and Fano effect individually and jointly. The characteristics of various Raman line shapes have been studied in terms of the broadening and asymmetry of Raman line shapes. It is shown that the asymmetry in the Raman line-shape function caused by these two effects individually does not add linearly to give asymmetry of line-shape generated by considering the combined effect. This indicates existence of interplay between the two effects.  The origin of interplay lies in the fact that Fano effect itself depends on quantum confinement effect and in turn provides an asymmetry. This can not be explained by considering the two effects’ contribution independent of each other.

\section{Introduction}
Raman Spectroscopy is widely considered as one of the most important and fast techniques to investigate vibrational property of materials[1–4]. Raman spectroscopy can also be used to predict the crystalline, amorphous or nanocrystalline nature of materials[1,5]. Raman scattering[6,7] is an inelastic scattering of incident photon and vibration / phonon of the materials. Due to momentum conservation law, in the crystalline bulk material, only zone centred (k=0) phonons participate in Raman scattering. As a result, symmetric, sharp peak is observed in the case of crystalline material centered at frequency corresponding to the zone centered phonon frequency. The equation of Raman line shape corresponds to the crystalline bulk material can be written as

\begin{equation}
I(\omega)=\dfrac{1}{[\omega-\omega_{0} ]^2+(\gamma/2)^2}
\end{equation}

Where, $\gamma$ is the full width half maximum (FWHM ) of Raman spectrum of the bulk material and $\omega_{0}$ corresponds to the zone centered optic phonon frequency. As an example, symmetric Raman peak in the case of bulk crystalline is observed at $~521 cm^{-1}$ with FWHM of $4 cm^{-1}$. In the case of nanostructures (NS), quantum confined (originating due to low dimensionality of the crystals) plays an important role and  k=0 selection rule gets relaxed. As a result a broaden red shifted asymmetric Raman spectrum is observed for quantum confined systems[4]. Apart from the confinement, there are other parameters like temperature , stress, electron-phonon interaction etc which induces additional changes in Raman line shape[8–11] .To understand the origin of asymmetric Raman line shape from confined systems, Richter et al[12]. have proposed a phonon confinement model (PCM), further it was  modified by Campbell et al[13] to take into account different size and shape of NS. According to PCM Raman line shape equation can be written as 

\begin{equation}
I_{L}(\omega)=\int_{0}^{1}\dfrac{e^{-\frac{k^2L^2}{4a^2}}d^{n}k}{[\omega-\omega_{0} ]^2+(\gamma/2)^2}
\end{equation}

Where \textit{k} is reduced wave vector, ‘\textit{a}’ being lattice parameter of material, ‘\textit{L}’ denotes the size of NS present in the sample. $\gamma$ is the FWHM of Raman spectrum of the bulk material. ‘\textit{n}’ being the order of confinement. This PCM has been successfully used in literature[14–16] to estimate the size and shape of nano materials. Equation 2 above can be derived qualitatively from Eq.1 by logically replacing its various terms by considering the underlying physics[4].  Confinement effect in the nanostructure is very much important to investigate the optical and electronic properties of material. In order to understand the actual effect of size (confinement effect) on the material properties, one should understand the characteristics of a Raman line-shape  and its responsive behavior as a result of an additional perturbation like electron-phonon interaction, stress, etc.

It is well accepted that in addition to quantum confinement effect, electron-phonon interaction or Fano interaction [14,17–19] also results broadening and asymmetry in Raman line shape of semiconductors. Basically Fano interaction is the interference of the continuum states and discrete levels lying in between the continuum [14,19]. Such type of Fano interaction has been studied in different type of system like atoms [20], molecules [21], and solids [22,23]. Fano effect in the semiconductor due to the inference between electronic continuum and discrete phonon results in an asymmetric Raman line-shape. The presence of electronic continuum in semiconductors is necessary (achieved by heavy doping) for observation of asymmetric Raman-Fano spectrum. In NS, a quasi- continuum of electronic states is available due to quantum confinement which can interfere with the discrete phonon[14,19]. On the other hand in heavily doped or in degenerate semiconductor the presence of electronic continuum is due to intra sub band excitations because of Fermi level being pushed as a result of heavily doping[18,24,25]. In the absence of quantum confinement effect,the Raman- Fano line-shape originating due to Fano interaction alone is given by following equation

\begin{equation}
I_{F}(\omega)=\left[\dfrac{(q+\varepsilon)^2}{1+\varepsilon^2}\right]
\end{equation}

Where $\varepsilon=\dfrac{\omega-\omega_{0}}{\gamma/2}$  ; $\gamma$ and $\omega_{0}$  being FWHM, and observed wavenumber of Fano transition respectively (reflected as peak position in Raman-Fano line-shape). Here the quantity ‘\textit{q}’ is known as Fano asymmetry parameter which measures the extent of Fano interference.  A low value of $|q|$ means strong Fano interaction and is inversely proportional to free charge carriers in the semiconductors[26].  In addition to asymmetry, a minimum in the line shape is also observed which is known as anti resonance and is a characteristic of the presence of Fano interaction in the system.

As discussed above, the Fano and confinement (F-C) effects are very important to explore the physics behind the phenomenon happening at nanoscale in the NS. The electronic transport and optical properties depend on size of NS and coupling in various transitions like Fano transition[19,23,27,28]. It is common between Fano and quantum confinement effects that both induces asymmetry in the Raman line-shape however there exists a typical difference in the nature of asymmetry as described below. Quantum confinement induced asymmetry in Raman line-shape is independent of doping type and a broader half width is observed in the lower wavenumber side of the Raman peak. On the other hand, asymmetry due to Fano interaction is doping-type dependent and asymmetry similar to quantum confinement effect is observed from an n-type semiconductor. The nature of Fano effect induced asymmetry in p-type semiconductor is opposite to that from n-type that is a broader half width is observed on the higher side of the peak position as compared to the lower wavenumber side. From the above discussion, it is obvious that the analysis of Raman line-shape is not very easy if both types of interactions are contributing to the Raman line-shape, especially for the semiconductors. Next challenge is to combine these effects for quatification of the individual effects. For better understanding, we have tried to present here an analytical description of both phenomenona distinctly and interplay between these.

\section{Analysis and discussion}
As discussed in the above section a symmetric Raman spectrum is observed for crystalline material having peak corresponding to the zone centre optic phonon. Fig. 1(a) shows a typical representation of symmetric Raman line shape showing peak corresponding to the zone frequency given in dispersion curve fig. 1(b) for Si. In nanostructured materials, phonons other than zone centre also contribute in the Raman scattering and results in broadening in the Raman line shape.

\begin{figure}[h]
    \centering
    \includegraphics[width=0.7\textwidth]{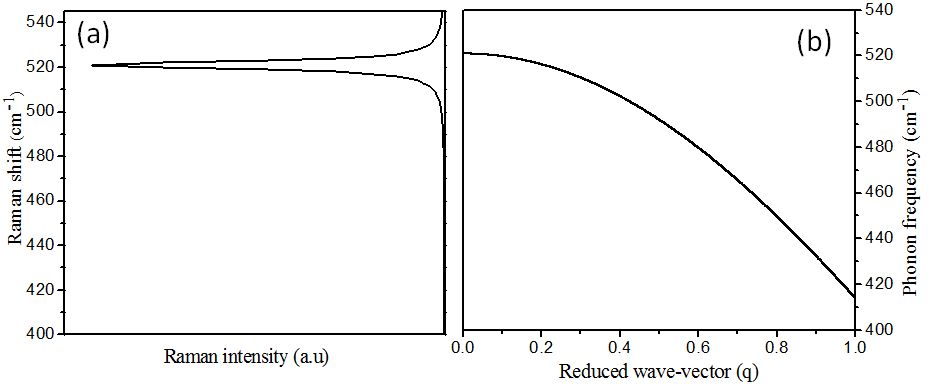}
    \caption{(a) Symmetric Raman line shape for C-Si (b) dispersion curve for Si}
    \label{fig:mesh1}
\end{figure}

 Figure 2(a) shows theoretical Raman line shape for Si NS obtained using Eq.2 by putting different crystallite sizes for a nanosystem which is confined in one dimension only (n=1) . It is evident from Fig. 2 that  with NS size decreasing, the broadening and asymmetry of line shape is increasing. This can be understood as follows, as the size of NS decrease, the range of phonons (corresponding to non-zero wavevectors (k), taking part in Raman scattering increase which results in non-zero intensities beyond peak frequency. This results in asymmetry and broadening in the Raman line shape. Therefore broadening and asymmetry of Raman line of any particular material is important parameter to analyze the material properties. We define asymmetric ratio as $AR=\dfrac{\gamma_{a}}{\gamma_{b}}$   , where $\gamma_{a}$  and $\gamma_{b}$  are the half widths on the low and high energy side of maximum. For consistency this definition will be used throughout this paper. Asymmetric ratio $AR_{L}$ as a function of nanostructure size is shown in figure 2(b) showing an increasing asymmetry with decreasing NS’s size.

 \begin{figure}[h]
    \centering
    \includegraphics[width=0.7\textwidth]{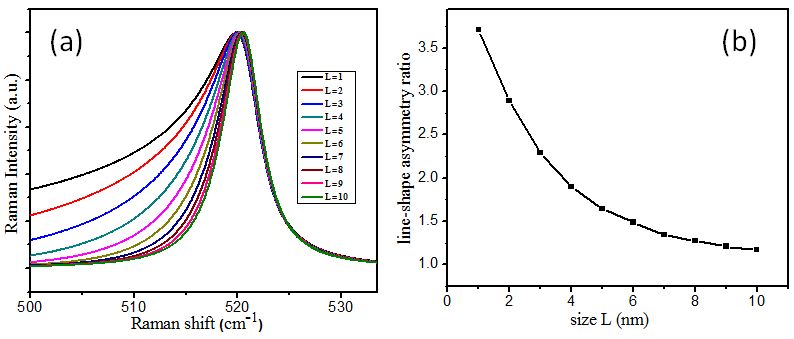}
    \caption{(a) Raman line-shapes obtained using Eq. 2 for different nanocrystallite sizes (in nanometers) (b) variation of asymmetry ratio as a function of nanostructures’size.}
    \label{fig:mesh1}
\end{figure}

 As discussed above, in addition to the confinement effect, Fano interference alone can also result in a broad and asymmetric Raman line shape. Figure 3 shows the Raman line shapes for different ‘\textit{q}’ values obtained using Eq. 3 where quantum confinement effect is not present. Figure 3(a) and Fig. 3(b) shows the Raman line shapes for negative and positive ‘\textit{q}’ values respectively. For the negative value of q the half width is more in the lower energy side of maximum whereas the half width is more in the higher energy side of the peak for the line-shape generated using positive q value. As a consequence of this, $AR>1$ is observed for negative q values whereas $AR<1$ is observed for positive q values. It is important here to mention that a line-shape with asymmetry ratio closer to one represents a symmetry line-shape whereas a line-shape having asymmetry ratio value away from one (towards zero or infinity) represents as asymmetry line-shape. 
Another important observation in Fig.3 is the presence of the anti resonance (minimum in line-shape) in the opposite direction of wider half-width side (higher energy side for negative q and lower energy side for positive value of q) which are the signatures of Fano interference [11,23,29]. Here it is clear from Fig. 3 that the broadening and asymmetry increases with the decreasing the absolute $|q|$ value. This can be seen more clearly in Fig. 3(c) which shows the variation of asymmetry ratio as a function of Fano asymmetry ratio ‘q’ It is clear from figure 3(c) that for the both negative and positive values of q the asymmetry ratio asymptotically approaches the value of one with increasing (decreasing) value of ‘$|q|$’ (1/$|q|$) meaning that a symmetrical line-shape will be obtained for large values of $|q|$. It is important here to mention here that asymmetry ratios vary between 1 and infinity in the negative q regime whereas it varies from one and zero in the positive q regime as shown in Fig. 3(c). Such an uneven variation in asymmetry ratio as a function of q results in non-symmetric AR v/s q plot in Fig. 3(c).    
From the above discussion, it is clear that Fano and confinement effects both induce asymmetry and broadening to the Raman line shape individually if the other is not present. We now consider the situation where both the effects (Fano and quantum confinement) are present simultaneous and contribute in the Raman line-shape as is the case with low dimensional systems with very high doping [23,26,29–34]. Therefore it seems to be very important to investigate the combined effect of confinement and Fano in nano materials.

\begin{figure}[h]
    \centering
    \includegraphics[width=0.7\textwidth]{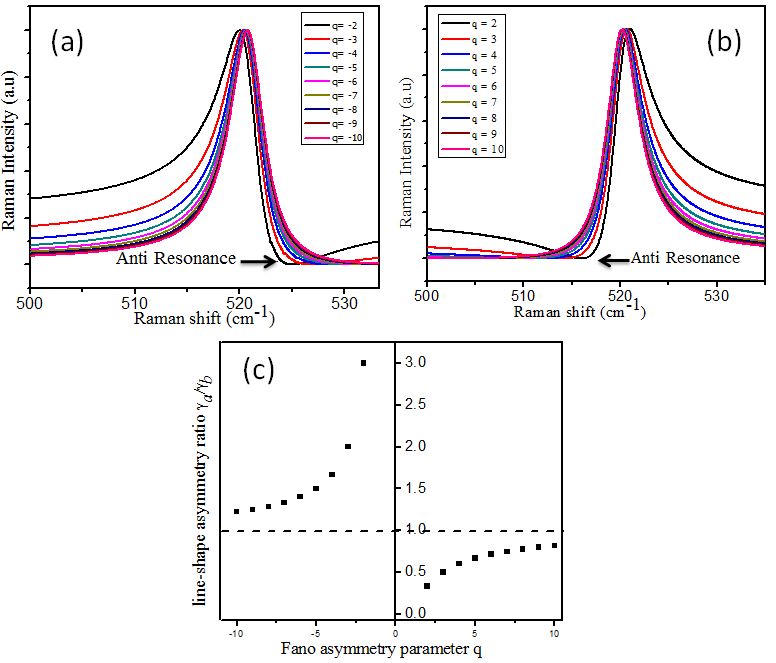}
    \caption{Raman spectra described by Eq 3. for different value of q  (a) for negative q values (b) for positive q values (c) Asymmetry ratio $AR_{q}$  for  positive and negative q value.}
    \label{fig:mesh1}
\end{figure}

Now to see the combined effect of F-C on Raman line shape the following equation has been taken which is already used by many authors in the literatures[14,15,35–37].

\begin{equation}
I_{L}(\omega)=\int_{0}^{1}e^{\dfrac{-k^{2}L^{2}}{4a^{2}}}\dfrac{(q+{\varepsilon}')^2}{1+{\varepsilon}'^2}d^{n}k
\end{equation}

Where ${\varepsilon}'=\dfrac{[\omega-\omega(k)]^2}{(\gamma/2)^2}$. Equation 4 takes care of both the effects Figure 4 (a) displays the Raman spectra having combined F-C effect for various value of negative q and fixed \textit{L} values 3 nm. These line-shapes show asymmetry, broadening and anti resonance together which are the signature of F-C effect. Figure 4(b) shows the asymmetry ratio as a function of Fano parameter for different L values. Figure 4(b) shows a general trend of increasing asymmetry ratio with decreasing q for a given crystallite size however the variation is rapid if the size is small. In other words the AR v/s q plot will indicate only Fano contribution for very high values of  \textit{L} where quantum confinement is negligible. This has already been reported in the literature[15] for the experimental work where the author claims that Fano interference is more pronounced for the smaller size of NS which supports the present results. One may interpret this is happening due to combined effect. But it is very interesting and important to mention here that the asymmetry ratio due to combined effect $(AR_{qL})$ is more than the sum of asymmetry ratios of $AR_{L}$ and $AR_{qL}$ asymmetry ratio if the two effects are present individually. That can be written as
                                                    
\begin{equation}
AR_{qL}>AR_{L}+AR_{q}                                                              
\end{equation}

This shows the interplay between Fano and confinement effects. 

\begin{figure}[t]
    \centering
    \includegraphics[width=0.7\textwidth]{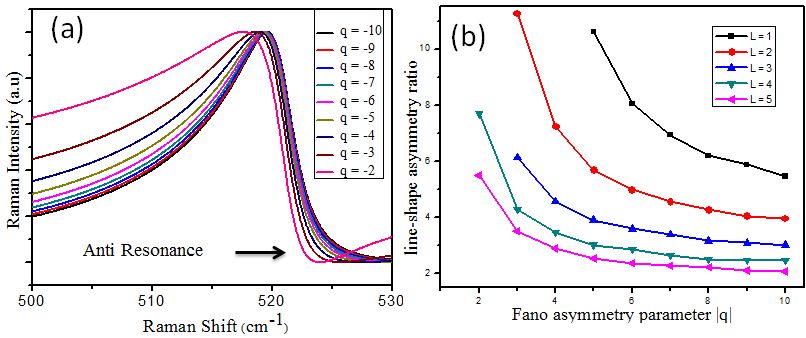}
    \caption{(a) Raman spectra described by Eq 4. for different value of q and fixed value of L=3 (b) asymmetry ratio as a function of q (Fano parameter) and  L (size of nanostructure in nanometers)}
    \label{fig:mesh1}
\end{figure}

\begin{table}[t]
\centering
\caption{Asymmetry ratio v/s size of nanostructures as obtained from Fig. 2}
\label{my-label}
\begin{tabular}{|c|c|}
\hline
{\bf Size of nanostructures(in nm)} & {\bf Asymmetry ratio} \\ \hline
1                                            & 3.71                  \\ \hline
\color[rgb]{0,0,1}{2}                                            & \color[rgb]{0,0,1}{2.89}                  \\ \hline
3                                            & 2.29                  \\ \hline
4                                            & 1.90                  \\ \hline
5                                            & 1.65                  \\ \hline
6                                            & 1.57                  \\ \hline
7                                            & 1.35                  \\ \hline
8                                            & 1.27                  \\ \hline
9                                            & 1.20                  \\ \hline
10                                           & 1.17                  \\ \hline
\end{tabular}
\end{table}

\begin{table}[]
\centering
\caption{Asymmetry ratio v/s  Fano parameter as obtained from Fig. 3 }
\label{my-label}
\begin{tabular}{|c|c|c|c|}
\hline
Value of Fano parameter q & Asymmetry ratio & Value of Fano parameter & Asymmetry ratio \\ \hline
-2                        & 3.00            & 2                       & 0.33            \\ \hline
\color[rgb]{0,0,1}{-3}                        & \color[rgb]{0,0,1}{2.00}            & 3                       & 0.50            \\ \hline
-4                        & 1.67            & 4                       & 0.60            \\ \hline
-5                        & 1.50            & 5                       & 0.67            \\ \hline
-6                        & 1.40            & 6                       & 0.71            \\ \hline
-7                        & 1.33            & 7                       & 0.75            \\ \hline
-8                        & 1.29            & 8                       & 0.78            \\ \hline
-9                        & 1.25            & 9                       & 0.80            \\ \hline
-10                       & 1.22            & 10                      & 0.82            \\ \hline
\end{tabular}
\end{table}

\begin{table}[]
\centering
\caption{ Asymmetry ratio v/s  Fano parameter q for fixed value  of  L (L=2 nm) as obtained from Fig.4}
\label{my-label}
\begin{tabular}{|c|c|}
\hline
{\bf Value of Fano parameter q} & {\bf Asymmetry ratio} \\ \hline
\color[rgb]{0,0,1}{-3}                                            & \color[rgb]{0,0,1}{11.25}                  \\ \hline
-4                                            & 7.24                  \\ \hline
-5                                            & 5.69                  \\ \hline
-6                                            & 4.98                 \\ \hline
-7                                            & 4.55                  \\ \hline
-8                                            & 4.27                  \\ \hline
-9                                            & 4.05                  \\ \hline
-10                                           & 3.95                  \\ \hline
\end{tabular}
\end{table}

\begin{table}[]
\centering
\caption{Asymmetry ratio v/s  Fano parameter q for fixed value  of  L (L=3 nm) as obtained from line-shape obtained using Eq. 4}
\label{my-label}
\begin{tabular}{|c|c|}
\hline
{\bf Value of Fano parameter q} & {\bf Asymmetry ratio} \\ \hline
\color[rgb]{0,0,1}{-3}                                            & \color[rgb]{0,0,1}{6.13}                  \\ \hline
-4                                            & 4.57                  \\ \hline
-5                                            & 3.90                  \\ \hline
-6                                            & 3.59                 \\ \hline
-7                                            & 3.38                  \\ \hline
-8                                            & 3.16                  \\ \hline
-9                                            & 3.09                  \\ \hline
-10                                           & 3.01                  \\ \hline
\end{tabular}
\end{table}

Table 1 and Table 2 show the asymmetry ratios ($AR_{L}$ and $AR_{q}$) due to individual effect of Confinement (L dependent) and Fano (q dependent).  The values in table 1 and table 2 are obtained using Eq. (2) and Eq.(3) respectively.  Table 3 and table 4 containing the asymmetry ratios due to the combine effect (F-C)  where the values are obtained using Eq. (4). One can see from table 1 to 4 that the value of $AR_{L}$ for $L= 2$ nm is 2.89 and the value of $AR_{q}$ for $q = -3$ is 2 giving $AR_{L}+AR_{q}=4.89$ but from table 3 the value of $AR_{qL}$ is 11.27. This has been highlighted in tables 1 to 3. The same calculation with the combination of L = 3 nm and $q= -3$  shows the value of $AR_{L}$ for L= 3 nm is 2.29 and the value of $AR_{q}$ for $q = -3$ is 2 so $AR_L+AR_q=4.29$  from table 4 the value of $AR_{qL}$ is 6.13. Validity of Eq.5 can be checked for other combinations of q and L using values in tables 1 to 4.Another important thing is that the contribution of Fano parameter $q=- 3$ on $L = 2$ nm NS induces the extra asymmetry of 6.38(11.27-4.89) whereas contribution of Fano parameter $q=- 3$ on $L = 3$ nm NS arises the extra asymmetry of 1.84 (6.13-4.29). In other words the deviation is more for smaller size which is originating due to the fact that Fano effect itself depends on the quantum confinement effect as reported earlier[15] as a result there exists an interplay between the two effects. The additional deviation in asymmetry ratio is more for $L=2$ (6.35) than $L=3$ (1.83) so one can say that Fano is more pounced for the smaller NS. The physics behind the inequality defined by Eq. (5) in the asymmetry ratios may be understood as the Fano effect is more pronounced for quantum confined system where the more possibilities of interference as there is more population of electronic continuum and more electron-phonon coupling is possible[36,37] .

\section{Conclusions}
Theoretical Raman line-shape functions for representing Fano effects and quantum confinement effects, in a semiconductor, show asymmetric nature of the line-shapes. The asymmetry of the line-shape due to quantum confinement depends only on size of the nanocrystalite and is independent of the doping type for a semiconductor. The asymmetry originating due to Fano effect alone depends on Fano asymmetry parameter and doping type in the semiconductor, which is indicated by sign of Fano parameter. In addition the asymmetric line-shape is accompanied by antiresonance which is otherwise absent if the asymmetry is due to quantum confinement effect. The Raman line-shape taking care of the combined effects of these two is also asymmetric in nature but the asymmetry is not a linear combination of the two effects when considered individually. The asymmetry in the line-shape generated due to combined effect is not equal to the summation of asymmetries coming due to individual effects. The inequality in asymmetry depends and size of the nanocrystallite suggesting a pronounced Fano effect for the smaller nanocrystallite size due to interplay between the two effects.

\section*{Acknowledgement}
Authors acknowledge financial support from Department of Science and Technology (DST), Govt. of India. Authors are also thankful to MHRD for providing fellowship.

\end{document}